\documentclass[aps,prl,pdflatex,twocolumn,notitlepage,superscriptaddress]{revtex4-2}
\usepackage[compat=1.0.0]{tikz-feynman}
\pdfoutput=1

\usepackage{amssymb}
\usepackage{multirow}
\usepackage{bm}
\usepackage{amsmath}
\usepackage{graphicx}
\usepackage{epstopdf}
\usepackage{subfigure}
\usepackage{natbib}
\usepackage{comment}
\usepackage{epsfig}
\usepackage{amsfonts}
\usepackage{mathrsfs}
\usepackage[toc,page,title,titletoc,header]{appendix}
\usepackage[colorlinks,linkcolor=blue,citecolor=blue,anchorcolor=blue]{hyperref}
\usepackage{cleveref}
\crefname{appendix}{App.}{Apps.}
\crefname{equation}{Eq.}{Eqs.}
\crefname{figure}{Fig.}{Figs.}
\crefname{table}{Tab.}{Tabs.}
\crefname{section}{Sec.}{Secs.}

\usepackage{braket}

\usepackage{dsfont,amsthm,amsbsy}

\def\Tr{{\rm Tr}}

\usepackage{fancyhdr}
\usepackage{xcolor}
\usepackage[normalem]{ulem}
\usepackage{bbm}
\usepackage{xcolor}

\begin{document}

\title{Establishing N\'eel order in the square-lattice antiferromagnetic Heisenberg model: a bootstrap proof}

\author{Nisarg Chadha}
\affiliation{Department of Physics, Harvard University, Cambridge MA 02138, USA}

\author{Taige Wang}
\affiliation{Department of Physics, Harvard University, Cambridge MA 02138, USA}
\affiliation{Materials Research Laboratory, Massachusetts Institute of Technology, Cambridge, MA 02139, USA}

\author{Pavel A. Nosov}
\affiliation{Department of Physics, Harvard University, Cambridge MA 02138, USA}

\author{Michael G. Scheer}
\affiliation{Department of Physics, Harvard University, Cambridge MA 02138, USA}

\author{Eslam Khalaf}
\affiliation{Department of Physics, Harvard University, Cambridge MA 02138, USA}

\begin{abstract}
Rigorously establishing spontaneous symmetry breaking (SSB) in the thermodynamic limit is generally challenging. This is exemplified by the two-dimensional (2D) $S = 1/2$ antiferromagnetic Heisenberg model where the existence of N\'eel order is not rigorously established despite overwhelming evidence from a variety of methods. We show that a bootstrap approach can be used to rigorously establish SSB and use it to prove the existence of N\'eel order in the 2D $S = 1/2$ XXZ model by deriving a strictly positive lower bound on the N\'eel order parameter for a set of discrete values of the anisotropy parameter $\Delta$ in the range $[0,2]$ including the isotropic (Heisenberg) point $\Delta = 1$. Our approach employs a generalization of the Kennedy--Lieb--Shastry inequalities, which use an infrared bound to reduce the existence of long-range order to bounds on certain finite-distance correlators. We compute these bounds for all sufficiently large systems with periodic boundary conditions using the bootstrap. An important ingredient in our proof is a novel method for obtaining exactly certified variational upper bounds on the energy density in the thermodynamic limit from matrix product states. Our approach can be readily generalized to other local lattice models for which infrared susceptibility bounds can be obtained.
\end{abstract}

\maketitle

\textit{Introduction}---Determining phase diagrams for microscopic quantum Hamiltonians is a central goal in many-body physics. For spontaneous symmetry breaking (SSB), the central challenge is to establish order in the thermodynamic limit \cite{Tasaki2020}. Unbiased numerical methods such as quantum Monte Carlo and exact diagonalization provide powerful evidence for SSB but rely on finite-size scaling \cite{Sandvik2010} and cannot directly access the thermodynamic limit. 
On the other hand, variational methods can access much larger or even infinite systems \cite{Jordan2008,Orus2014} but cannot by themselves certify long-range order.

The absence of a method to certify long-range order means that even for models where the existence of such order is physically well established based on experimental, theoretical, and numerical evidence, a rigorous proof can be absent. This is exemplified by the square-lattice antiferromagnetic Heisenberg model. Pioneering work by Dyson, Lieb, and Simon \cite{dyson1978phase} and subsequent developments \cite{neves1986long}, particularly by Kennedy, Lieb, and Shastry (KLS) \cite{kennedy1988existence}, reduced the problem of the existence of long-range N\'eel order in the model to the violation of certain inequalities involving local correlators. Together with variational energy bounds, the violation of these inequalities rigorously established N\'eel order for spin $S\geq1$ \cite{neves1986long,kennedy1988existence} and for spin-$1/2$ with sufficiently strong easy-axis or easy-plane anisotropy \cite{PhysRevLett.61.2585,PhysRevLett.61.2582,Nishimori1989,ozeki1989long,wischmann1991extended,you2009long}. However, the most interesting case, the isotropic square-lattice Heisenberg model with $S = 1/2$, has remained an open problem \cite{LiebProblem1999}. Numerical evidence that the KLS inequalities are violated for the isotropic spin-$1/2$ model, once sufficiently long-range correlators are included, was obtained by L\"ow using finite-size quantum Monte Carlo \cite{PhysRevB.76.220409}.

In this Letter, we show that bootstrap methods \cite{Navascues2007,Navascues2008,Barthel2012,Baumgratz2012,Han2020,Lin2020,Cho2022,Hastings2022,Kazakov2023,Mazziotti2023,Berenstein2024,Wang2024,Fawzi2024,Gao2025,Cho2025,Schouten2025,Scheer2025,Scheer2026a,Scheer2026b,Cho2026a,Cho2026b,Gao2026a,Paul2026,Chadha2026,Onder2026,Cho2026c} can be used to establish SSB through finite computations yielding rigorous lower bounds on the order parameter directly in the thermodynamic limit. We use this method to prove the existence of N\'eel order in the spin-$1/2$ square-lattice Heisenberg antiferromagnet at zero temperature by obtaining a strictly positive lower bound on the N\'eel order parameter. This implies the existence of ground states with spontaneous staggered magnetization in the thermodynamic limit \cite{Koma1993}. All numerical bounds are converted into exact rational certificates, which makes them rigorous and yields a full mathematical proof.

\begin{figure}[t]
\centering
\includegraphics[width=0.75\linewidth]{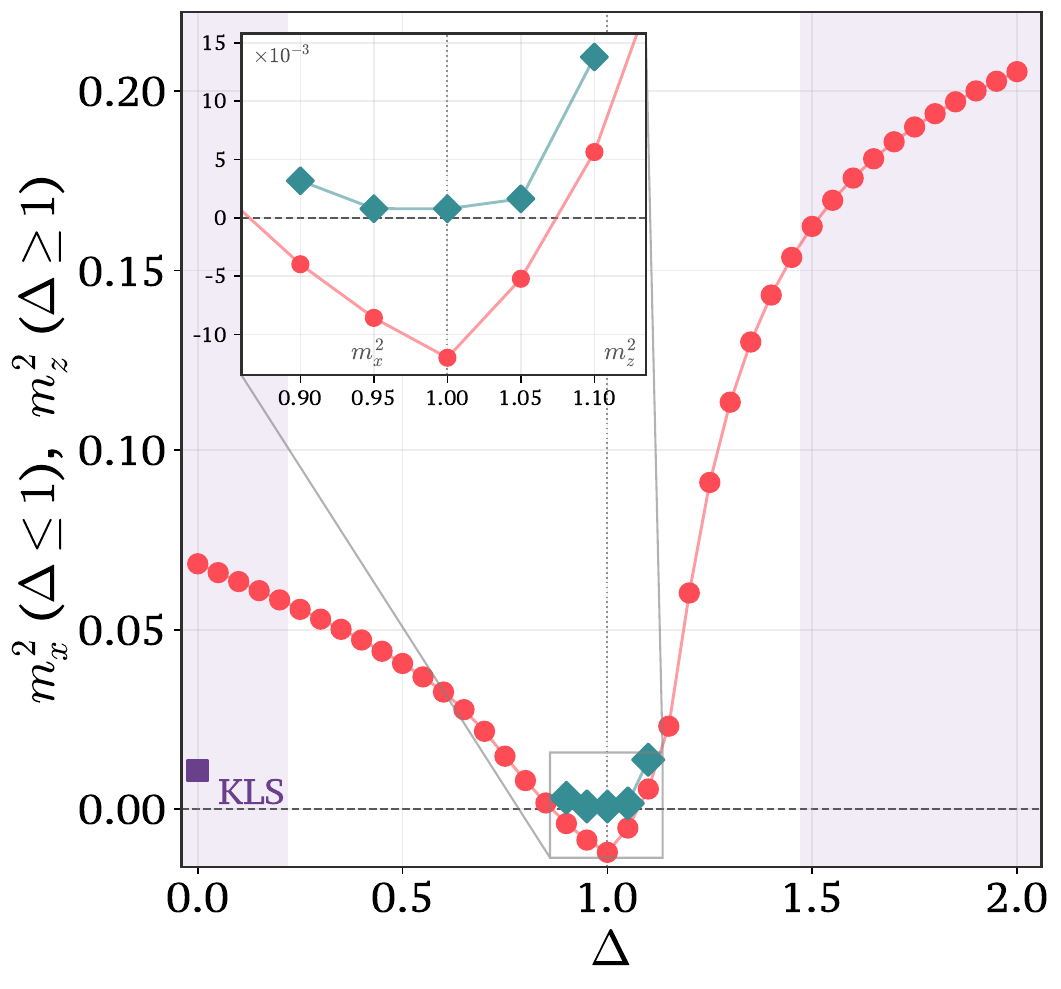}
\caption{\textbf{Exactly certified lower bounds on the order parameter.}
The figure shows lower bounds on $m_x^2$ for $\Delta\leq 1$ and $m_z^2$ for $\Delta \geq 1$ as a function of $\Delta$.
Red circles denote the bounds using a smaller operator set with exactly certified tensor-network energy ceilings.
The teal diamonds show the lower bounds for $\Delta$ close to 1 using larger operator sets and tighter variational ceilings using the method described in the text.
The shaded columns mark the ranges $0\le\Delta\le0.22$ and $\Delta\ge1.47$ where N\'eel order was
previously proven~\cite{wischmann1991extended}. 
The purple square shows the KLS bound at $\Delta = 0$~\cite{PhysRevLett.61.2582}.
}
\label{fig:lowerbounds}
\end{figure}

Our proof builds on the KLS inequalities, which use infrared bounds to analytically control long-wavelength fluctuations and reduce the existence of long-range order to bounds on finite-distance correlators. Our key advances are (i) the derivation of a tighter family of KLS-type inequalities with optimizable parameters, (ii) the use of the bootstrap to obtain rigorous bounds on the correlators that enter such inequalities, and (iii) a new approach to obtain certified bounds on the energy density in the thermodynamic limit from matrix product states (MPSs). The latter is used to tighten the bootstrap bounds, leading to a strictly positive lower bound on the N\'eel order parameter for a set of discrete values of the anisotropy parameter $\Delta$ in the range $[0,2]$, including the isotropic point $\Delta = 1$ (see \cref{fig:lowerbounds}). In principle, a positive lower bound at a given $\Delta$ also extends to a finite interval around $\Delta$ (see End Matter).

We emphasize that the main structural property of the model that was needed, beyond being a local lattice model, is the infrared bound which is usually a property of models with reflection positivity \cite{frohlich1976infrared,dyson1978phase}. More broadly, this framework opens a route to the rigorous and systematic determination of phase diagrams in interacting quantum systems.

\textit{The XXZ Model}--- We consider the spin-$1/2$ XXZ model on a square-lattice torus $\Lambda$ with unit lattice spacing:
\begin{equation}
\label{eq:ham}
    H_\Delta =\sum_{\langle ij\rangle}h_{ij} =  \sum_{\braket{ij}} (X_iX_j + Y_iY_j + \Delta Z_iZ_j),
\end{equation}
where $S = X, Y, Z$ are the spin-1/2 operators given by $S_i = \sigma^S_i/2$ in terms of the Pauli matrices.
$H_{\Delta}$ has a $U(1)$ internal symmetry with respect to rotations about the $z$-axis, and a $\mathbb{Z}_2$ Ising spin flip symmetry that flips $Z$ and $Y$ while leaving $X$ invariant.
The model reduces to the $XY$ model when $\Delta = 0$, and to the Ising model when $\Delta \rightarrow \infty$.
$\Delta = 1$ corresponds to the $\text{SU}(2)$-symmetric Heisenberg model.

We define the thermodynamic ground state as the limit of the finite-volume ground states on $\Lambda$ as $|\Lambda|\rightarrow\infty$.
For every finite $\Lambda$, there is a unique, fully symmetric ground state \cite{lieb1962ordering} with correlations denoted as $\braket{\cdot}_{\Lambda}$.
Although the Mermin--Wagner theorem \cite{PhysRevLett.17.1133} precludes order at any finite temperature, the thermodynamic ground state can spontaneously break symmetries \cite{neves1986long}, evidenced by a singularity in the thermodynamic structure factor $S(\mathbf{q}) = \lim_{|\Lambda|\rightarrow\infty} S_{\Lambda}(\mathbf{q})$, where
\begin{equation}
    S_{\Lambda}(\mathbf{q}) = \braket{\tilde{X}_{\mathbf{q}} \tilde{X}_{-\mathbf{q}}}_{\Lambda},
\end{equation}
and the Fourier transform is defined as $\tilde{X}_{\mathbf{q}}=\frac{1}{\sqrt{|\Lambda|}} \sum_{\mathbf{r}} e^{i\mathbf{q} \cdot \mathbf{r}} X_{\mathbf{r}}$. 

The N\'eel order parameter is given by the staggered magnetization $M_x=\sum_{\mathbf{r}} \eta_{\mathbf{r}} X_{\mathbf{r}}$, where $\eta_{\mathbf{r}} = \pm 1$ on the A and B sublattices respectively. The squared order parameter density is defined by
\begin{equation}
  m_{\Lambda,x}^2 = \frac{\langle M_x^2\rangle_\Lambda}{|\Lambda|^2} = \frac{S_\Lambda(\mathbf{Q})}{|\Lambda|} \text{ where } \mathbf{Q} = (\pi,\pi).\label{eq:m_x}
\end{equation}
In the thermodynamic limit, $m_x^2=\lim_{|\Lambda|\rightarrow\infty}m_{\Lambda,x}^2$ measures the coefficient of a $\delta$-function singularity in the structure factor at $\mathbf{q} =\mathbf{Q}$.

\textit{The KLS construction}---We start by reviewing the construction of Kennedy-Lieb-Shastry (KLS) \cite{PhysRevLett.61.2582}, which establishes a lower bound on the magnetization in terms of local correlators using an upper bound on $S(\mathbf{q})$ introduced by Refs. \cite{frohlich1976infrared, dyson1978phase} along with the sum rules obeyed by $S(\mathbf{q})$. The KLS inequality, combined with variational energy upper bounds, was used previously to establish N\'eel order in the 2D $XY$ model and in the Heisenberg model in more than two spatial dimensions. However, it was insufficient to establish N\'eel order in 2D Heisenberg model.

The KLS construction begins with the sum rule
\begin{equation}
\label{eq:sumrule}
    \sum_{\mathbf{q}} \frac{1}{|\Lambda|} \gamma_{\mathbf{n}}(\mathbf{q}) S_{\Lambda}(\mathbf{q}) = \braket{{X_{\mathbf{0}} X_{\mathbf{n}}}} = C^X_{\mathbf{n}}, \, \mathbf{n}\in\mathbb{Z}^2,
\end{equation}
where $\gamma_{\mathbf{n}}(\mathbf{q}) = \cos(n_xq_x+n_yq_y)$. For $\mathbf{n} = \mathbf{0} = (0,0)$, $C^X_{\mathbf{0}}=1/4$ for any state, whereas other choices of $\mathbf{n}$ relate various harmonics of the structure factor to local correlators. Although KLS only considered the cases of $\mathbf{n} = \mathbf{0}$ or $(1,0)$, \cref{eq:sumrule} represents a straightforward generalization.

Taking linear combinations of Eq.~\eqref{eq:sumrule} for various $\mathbf{n}$ and isolating the $\mathbf{q} = \mathbf{Q}$ term gives the generalized sum rule:
\begin{equation}
    \label{eq:gensumrule}
    w(\mathbf{Q})m_{\Lambda,x}^2  = \frac{a_0}{4} +\sum_{\mathbf{n}} a_{\mathbf{n}}C^X_{\mathbf{n}}-\frac{1}{|\Lambda|}\sum_{\mathbf{q}\neq \mathbf{Q}} w(\mathbf{q})S_{\Lambda}(\mathbf{q})
\end{equation}
where $w(\mathbf{q})$ is defined as $w(\mathbf{q}) = a_0 + \sum_{\mathbf{n}} a_{\mathbf{n}}\gamma_{\mathbf{n}}(\mathbf{q})$ and is normalized such that $w(\mathbf{Q}) = 1$. 
Since $S_{\Lambda}(\mathbf{q})\geq 0$, we obtain a lower bound on the order parameter by dropping the regions where $w(\mathbf{q}) <0$:
\begin{equation}\label{eq:nearest-neighbor-sum-rule}
    m_{\Lambda,x}^2 \geq \frac{a_0}{4} +\sum_{\mathbf{n}} a_{\mathbf{n}}C^X_{\mathbf{n}}-\frac{1}{|\Lambda|}\sum_{\mathbf{q}\neq \mathbf{Q}} \{w(\mathbf{q})\}_+S_{\Lambda}(\mathbf{q}),
\end{equation}
where $\{w(\mathbf{q})\}_+ = \max(0,w(\mathbf{q}))$.
The first two terms on the right-hand side depend only on local correlators, but the last one involves a sum over all momenta, which cannot be expressed in terms of finitely many local correlators.

To deal with the last term, KLS derived an upper bound on $S_{\Lambda}(\mathbf{q})$ using an infrared bound on the susceptibility as we review below. 
The dynamical structure factor is
\begin{equation}
  S_{\Lambda}(\mathbf{q},\omega)=\sum_n |\langle n|\tilde{X}_{\mathbf{q}}|0\rangle_{\Lambda}|^2\,\delta(\omega-\omega_n)\ \ge 0,
\end{equation}
where $\omega_n=E_n-E_0$. Some of its moments $m_p(\mathbf{q}) = \int_0^\infty d\omega\,\omega^p\,S_{\Lambda}(\mathbf{q},\omega)$ are familiar objects:
$m_0=S_{\Lambda}(\mathbf{q})$, $m_1=\tfrac12 D_x(\mathbf{q})$ and $m_{-1} = \tfrac12\chi_x(\mathbf{q})$, where $D_x(\mathbf{q})=\langle[\tilde{X}_{\mathbf q}^\dagger,[H,\tilde{X}_{\mathbf{q}}]]\rangle$ is the double commutator and $\chi_x(\mathbf{q})$ is the 
static susceptibility. 
The Cauchy--Schwarz inequality in $\omega$ gives
\begin{equation}
  S(\mathbf{q})\le\sqrt{m_1 m_{-1}}=\tfrac12\sqrt{D_x(\mathbf{q})\,\chi_x(\mathbf{q})} .
  \label{eq:CS}
\end{equation}
The double commutator only retains local terms. In particular, for the nearest-neighbor Hamiltonian, it only involves nearest-neighbor correlations,
\begin{equation}
    D_x(\mathbf{q}) = -4\big[(1-\Delta\gamma(\mathbf{q}))C^X_{1,0} +(\Delta-\gamma(\mathbf{q}))C^Z_{1,0}\big],
\end{equation}
where we have defined $\gamma(\mathbf{q}) = \frac{1}{2}(\gamma_{1,0}(\mathbf{q})+\gamma_{0,1}(\mathbf{q}))$. 
On the other hand,
$\chi_x(\mathbf{q}) =2\langle \tilde{X}_{\mathbf{q}}^\dagger (H-E_0)^{-1}\tilde{X}_{\mathbf{q}}\rangle$ depends on the whole low-energy spectrum. 
A small amount of spectral weight at very low frequency
changes every local expectation value negligibly but changes $\chi$ by an arbitrary amount. 
Thus, while \cref{eq:CS} gives us an upper bound, it simply trades one unknown, $S(\mathbf{q})$, for another, $\chi_x(\mathbf{q})$.

So far no special property of the Hamiltonian has been used apart from locality. Using the reflection positivity of the Hamiltonian, Refs.~\cite{frohlich1976infrared, dyson1978phase} derived the infrared bound:
\begin{equation}
  \chi_x(\mathbf{q})\le\bar{\chi}_x(\mathbf{q}) = \frac{1}{4(1+\gamma(\mathbf{q}))},\qquad \mathbf{q}\neq \mathbf{Q},
  \label{eq:IR}
\end{equation}
which holds in the Gibbs state at any temperature, and hence in the ground state, on any even torus and for every $\Delta\geq 0$
(see Ref.~\cite{bjornberg2022reflection} for the precise hypotheses, and the End Matter for a derivation).

In two dimensions at zero temperature, the singularity of this bound at $\mathbf{q}=\mathbf{Q}$ is integrable. Together, \cref{eq:CS,eq:IR} give an upper bound on the structure factor that only involves local correlators:
\begin{equation}
  S(\mathbf{q})\le U(\mathbf{q})=\frac{1}{2}\sqrt{-\frac{[(1-\Delta\gamma(\mathbf{q}))C^X_{1,0} + (\Delta-\gamma(\mathbf{q}))C^Z_{1,0}]}{1+\gamma(\mathbf{q})}}.
  \label{eq:KLScap}
\end{equation}
Inserting \cref{eq:KLScap} into \cref{eq:nearest-neighbor-sum-rule} produces the KLS inequality~\cite{PhysRevLett.61.2582}
\begin{equation}
    m_{\Lambda,x}^2\geq \frac{a_0}{4} +\sum_{\mathbf{n}}a_{\mathbf{n}}C^X_{\mathbf{n}} - \frac{1}{|\Lambda|}\sum_{\mathbf{q}\neq\mathbf{Q}} \{w(\mathbf{q})\}_+ U(\mathbf{q}).
\end{equation}
In the thermodynamic limit, the sum becomes an integral over the Brillouin zone.
KLS used variational upper bounds and the Kubo inequality $|C^Z_{1,0}|\leq-C^X_{1,0}$ to establish N\'eel order in the ground-state of the XY model.

\textit{Generalizing the KLS inequality}---The only non-local input entering the KLS argument is the infrared bound in \cref{eq:IR}; everything else is a statement about the moments of the non-negative function $S(\mathbf q,\omega)$. Thus, including higher moments may help us tighten the bound. 
Indeed, for inversion symmetric states, the odd moments $m_{2p+1}(\mathbf{q}) = m_{2p+1}(-\mathbf{q})=\frac{1}{2}\braket{[\tilde{X}_{\mathbf{q}}^{\dagger}, \mathrm{ad}_H^{2p+1} \tilde{X}_{\mathbf{q}}]}$ are equal to linear combinations of local correlators.
Each commutator with the Hamiltonian enlarges the support of an operator by at most one site, so $\mathrm{ad}_H^{2p+1}X_{\mathbf{0}}$ is a sum of Pauli strings on at most $2p+2$ sites within a taxicab distance $2p+1$ of the origin, and $\braket{[X_{\mathbf{r}}, \mathrm{ad}_H^{2p+1}X_{\mathbf{0}}]}=0$ unless $|\mathbf{r}|\leq 2p+1$.
Therefore, each $m_{2p+1}(\mathbf{q})$ is a finite trigonometric polynomial in $\mathbf{q}$ whose coefficients are linear combinations of local correlators:
\begin{equation}
  m_{2p+1}(\mathbf{q})=\sum_{|\mathbf{r}|\le 2p+1}\gamma_{\mathbf{r}}(\mathbf{q})\,G_{2p+1}(\mathbf{r}),
\end{equation}
where $G_p(\mathbf{r})=\tfrac12\langle[X_{\mathbf{r}},\mathrm{ad}_H^{p}X_{\mathbf 0}]\rangle$. 
The even moments, including $m_0(\mathbf{q})=S(\mathbf{q})$, are genuinely non-local and remain unknown. 

Bounding $S(\mathbf q)$ thus becomes a moment problem: how large can
$m_0=S(\mathbf q)$ be for a nonnegative function $S(\mathbf q,\omega)$ on
$\omega>0$ whose odd moments $m_1,m_3,\dots,m_{2J-1}$ are given and whose
moment $m_{-1}$ is at most $\bar\chi(\mathbf q)/2$?

Every polynomial 
\begin{equation}
\label{eq:poly}
  P(x)=\lambda-x+\sum_{p=1}^{J}\alpha_{2p-1}\,x^{2p}\ \ge0\quad\text{for all }x\ge0
\end{equation}
gives such an upper bound: since $\int d\omega \frac{P(\omega)}{\omega} S(\mathbf{q},\omega)\geq 0$, we get
\begin{equation}
  S(\mathbf q)\le f_P(\mathbf q)=\lambda\,\frac{\bar\chi(\mathbf q)}{2}
  +\sum_{p=1}^{J}\alpha_{2p-1}\,m_{2p-1}(\mathbf q).
  \label{eq:momentcap}
\end{equation}
Apart from the $-x$ term, $P$ contains only even powers, so dividing by $\omega$ and integrating against $S(\mathbf q,\omega)$ produces only $m_{-1}$, $m_0$ and the local odd moments.
Choosing the best $P$ gives the tightest cap, which equals the largest $m_0$ compatible with the
infrared bound and the given moments, i.e.\ the solution of a truncated Stieltjes
moment problem~\cite{kreuin1977markov,schmudgen2017moment}.
For $J=1$, $P\ge0$ if and only if $4\lambda\alpha_1\ge1$, and the optimum gives the KLS cap~\eqref{eq:KLScap}.

A different polynomial $P_{\mathbf q}$ may be used at each momentum. For any family $P=\{P_{\mathbf q}\}$ satisfying Eq.~\eqref{eq:poly}, inserting the cap into the generalized sum rule gives
\begin{equation}
\label{eq:neelbound}
  m_x^2\ \ge\ \frac{a_0}{4}+\sum_{\mathbf n}a_{\mathbf n}C^X_{\mathbf n}
  -\int\!\frac{d^2 \mathbf{q}}{(2\pi)^2}\,\{w(\mathbf q)\}_+\,f_P(\mathbf q).
\end{equation}
The tightest bound uses, at each $\mathbf q$, the polynomial that minimizes $f_P(\mathbf q)$; any other choice gives a larger cap and hence a weaker but valid bound.

\textit{Easy-axis order for $\Delta\geq 1$}---For $\Delta\geq1$ the spins order along $z$, and the same construction bounds $m_z^2=\lim_{\Lambda\to\infty}\frac{1}{|\Lambda|^2}\langle(\sum_{\mathbf{r}}\eta_{\mathbf{r}}Z_{\mathbf{r}})^2\rangle$, with $\tilde{Z}_{\mathbf{q}}$ instead of $\tilde{X}_{\mathbf{q}}$, the correlators $C^Z_{\mathbf{n}}$ appearing in the sum rule.
The corresponding infrared bound changes to
$\chi_z(\mathbf{q})\le1/[4\Delta(1+\gamma(\mathbf{q}))]$~\cite{dyson1978phase, bjornberg2022reflection} (End Matter), and the double commutator becomes
\begin{equation}
    D_z(\mathbf{q})=-8\,C_{1,0}^X(1-\gamma(\mathbf{q})).
\end{equation}
The higher moments follow from
$\mathrm{ad}_H^{2p+1}Z_{\mathbf 0}$ exactly as before.

\textit{Bootstrapping ground-state correlations}---\cref{eq:neelbound} uses a variety of local correlators beyond nearest-neighbor $XX$ and $ZZ$ correlations both through the inclusion of higher $\mathbf{n}$ in the sum rule and the consideration of higher moments in the $S(\mathbf{q})$ upper bound.
Previous approaches to this problem so far could only get rigorous one-sided bounds on nearest-neighbour correlations coming from the variational ceiling, with no information about the longer-range and multi-site
correlators. 
The full freedom in Eq.~\eqref{eq:neelbound} therefore could not be exploited, and earlier bounds were restricted to the nearest-neighbor sum rule and the KLS cap, allowing the proof of easy-plane order for $0\le\Delta\leq 0.22$ and easy-axis order for $\Delta\geq1.47$~\cite{PhysRevLett.61.2585,ozeki1989long,nishimori1989long,Nishimori1989,wischmann1991extended}.
However, the proof of order around the Heisenberg point $\Delta = 1$ has remained an open problem~\cite{tasaki2019long}.

We get rigorous control of arbitrary local ground-state correlations using the Hamiltonian bootstrap framework \cite{Scheer2026a}.
Instead of explicitly computing the ground-state, the bootstrap looks for values of correlators that are compatible with a set of consistency conditions defined for the ground-state. The ground-state expectation value $\langle\cdot\rangle$ is a linear map from local operators to the complex numbers, so it is fixed by its values on a basis of Pauli strings. We choose a finite set of local operators $O_a$ and let $\mathbf y$ collect the expectation values of the distinct Pauli strings that appear in the products $O_a^\dagger O_b$ and $O_a^\dagger[H,O_b]$; every expectation value we need is then a linear function of $\mathbf y$.
We use the symmetries of the Hamiltonian to identify Pauli strings related by lattice and internal symmetries.
Each consistency condition is either a linear equality or inequality in $\mathbf{y}$, or a linear matrix inequality $M(\mathbf{y})\succeq 0$, where the entries of $M$ depend linearly on $y$.
Because positive semidefinite (PSD) matrices form a convex cone, these conditions define a convex feasible set
$\mathcal K$, which is guaranteed to contain the true ground-state values.

The conditions we impose are
\begin{equation}
\begin{aligned}
  &\langle I\rangle=1,\quad \langle O_a^\dagger O_b\rangle\succeq0,\quad
   |\langle P\rangle|\le1,\\
  &\langle[H,O]\rangle=0,\quad \langle O_a^\dagger[H,O_b]\rangle\succeq0,
\end{aligned}
\label{eq:constraints}
\end{equation}
where $\succeq0$ means that the matrix with indices $a,b$ is positive semidefinite. 
The first line holds in any physical state: normalization,
positivity, and the norm bound $|\langle P\rangle|\le1$ on every Pauli string $P$.

The second line characterizes ground states of $H$: stationarity holds for any eigenstate, and perturbative positivity,
$\langle O^\dagger[H,O]\rangle=\langle O^\dagger(H-E_{gs})O\rangle\ge0$, states that no operator can lower the energy. 
These conditions involve only local operators and commutators with $H$, so they remain meaningful in the thermodynamic limit~\cite{Fawzi2024}.

We also impose the lattice and spin symmetries of the ground state, which is unique on every even torus, as equalities between expectation values of symmetry-related operators, and use them to block-diagonalize the PSD constraints~\cite{Scheer2026a}. 
All these conditions hold on every large enough even torus, and therefore in the infinite-volume limit.

While these relations are, in principle, sufficient to exactly define the ground states when applied to all operators in the quasi-local algebra, in practice, we can only apply them for a small subset of operators supported on a finite region. 
Thus, it is useful to supplement these with other relations that incorporate additional information or rely on a specific property of the Hamiltonian. 
For the XXZ model, translation and $U(1)$ symmetry make the two bonds per site contribute equally, so the ground-state energy density is
\begin{equation}
\epsilon_0\equiv\lim_{|\Lambda|\to\infty}\frac{E_0(\Lambda)}{|\Lambda|}
=4C^X_{1,0}+2\Delta C^Z_{1,0}.
\label{eq:energy_corr}
\end{equation}
A certified variational upper bound $\epsilon_0\le\varepsilon_{\rm var}$, which we obtain with tensor networks below, therefore becomes the linear constraint $4C^X_{1,0}+2\Delta C^Z_{1,0}\le\varepsilon_{\rm var}$ on $\mathbf y$.
Unlike the conditions above, this constraint need not hold on a finite torus, where $E_0(\Lambda)/|\Lambda|$ can lie below $\epsilon_0$, but it holds in the infinite-volume limit, which is all we use.
We also use reflection positivity (RP), $\langle A_a\,\Theta(A_b)\rangle\succeq0$, and the Marshall sign rule, $\eta_W\langle W\rangle\ge0$.
RP follows from the structure of the Hamiltonian: $\Theta$ maps operators $A_a$ on one side of a bond-centered mirror plane to their reflections $\Theta(A_a)$ on the other side (see End Matter).
The Marshall sign rule~\cite{marshall1955antiferromagnetism,lieb1962ordering} states that the ground-state expectation value of any product $W$ of raising and lowering operators $S^{\pm}$ has the sign $\eta_W = (-1)^{\#(\text{sites of }W\text{ on the }B\text{ sublattice})}$.

For a fixed choice of $w$ and $P$, the cap $f_P(\mathbf{q})$ is linear in $\mathbf{y}$, making the right-hand side of Eq.~\eqref{eq:neelbound} an affine function $F_{w,P}(\mathbf{y})$.
Since the true ground-state values $\mathbf{y}^\star$ lie in $\mathcal K$, a single semidefinite program bounds all the correlators in $F_{w,P}$ jointly:
\begin{equation}
  m_x^2\ \ge\ F_{w,P}(\mathbf y^\star)\ \ge\ \min_{\mathbf y\in\mathcal K}F_{w,P}(\mathbf y).
\end{equation}

Since $m_{2p+1}$ already includes a number of strings on up to $2p+2$ sites, getting tight bounds for each of these strings requires progressively larger operator sets.
For instance, $m_7$ alone requires about $1.9\times10^5$ distinct strings, with most of them supported on 8 sites.
In practice, we restrict ourselves to $J=3$, which includes odd moments up to $m_5$.

The optimal bound for $J=1$ is the KLS cap, which includes a square root over nearest-neighbour correlators.
However, with fixed coefficients, the $J=1$ polynomial in Eq.~\eqref{eq:poly} is a
tangent line to this square root with $\alpha_1=1/(4\lambda)$,
$\lambda\bar\chi/2+D/(8\lambda)\ge\tfrac12\sqrt{D\bar\chi}$ for every $\lambda>0$, with equality at one value of $\lambda$. Choosing $\lambda$ at a reference point therefore gives a valid cap that is affine in $\mathbf y$ and
tight at that point. 
We use the KLS cap close to $Q$, where $\bar\chi$ diverges, and the $J=3$ cap elsewhere (End Matter).

\textit{Optimizing the weights and the cap}---Every weight $w$ with $w(\mathbf{Q})=1$, and every set of cap polynomials
$P=\{P_q\}$ nonnegative on $[0,\infty)$, gives a valid bound, so
\begin{equation}
  m_x^2\ \ge\ \max_{w,\,P}\ \min_{\mathbf y\in\mathcal K}\ F_{w,P}(\mathbf y).
  \label{eq:maxmin}
\end{equation}
For fixed $P$, the maximization over $w$ can be combined with the bootstrap
into a single semidefinite program (End Matter). The polynomials are chosen
from the moment problem at each momentum. For given $m_1$, $m_3$, $m_5$ and
$\bar\chi$, the largest $S(\mathbf{q})$ is attained when $S(\mathbf q,\omega)$ consists of two
delta functions, a low-frequency magnon-like peak and a high-frequency
continuum peak. The optimal polynomial is the one that vanishes, with zero
slope, at exactly these two frequencies, so that the cap is saturated. We
compute these frequencies from the moments at the current optimum, round the
polynomial's coefficients to exact rationals, and verify exactly that it is
nonnegative on $[0,\infty)$. We then re-optimize $w$ with the new polynomials
and repeat; a few rounds suffice. A final semidefinite program at fixed $w$
and $P$ gives the reported bound. Its validity does not depend on how $w$ and
$P$ were found.

\textit{Certified variational upper bound}---Variational energy bounds have long played an important role in rigorous studies of antiferromagnetic order~\cite{PhysRevLett.61.2582,kennedy1988existence}. 
Through Eq.~\eqref{eq:energy_corr}, a rigorous upper bound on $\epsilon_0$ directly constrains the nearest-neighbor correlations entering the N\'eel bound. In fact, our bootstrap lower bound on the N\'eel order parameter at $\Delta=1$ becomes positive only after imposing the improved energy bound derived below.

\begin{figure}[t]
\centering
\includegraphics[width=\linewidth]{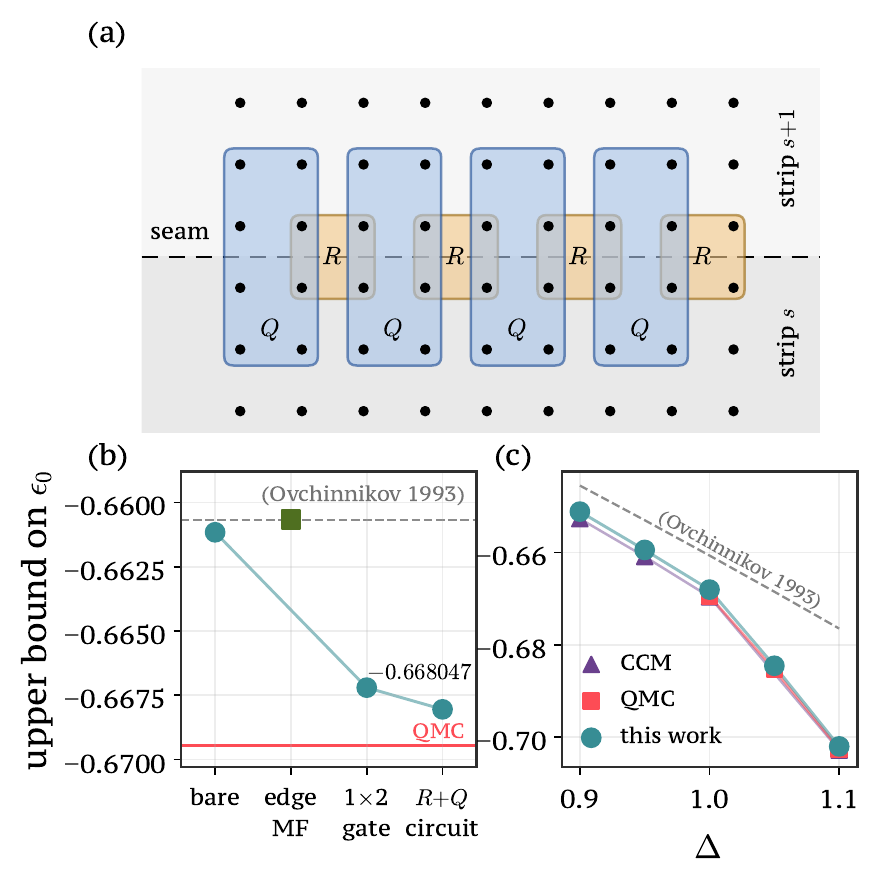}
\caption{\textbf{Certified variational upper bounds.}
(a) Infinite width-$W$ MPS strips are coupled across their seams by a shallow circuit of four-spin $R$ and eight-spin $Q$ gates.
(b) Certified hierarchy at $\Delta=1$ for $W=12$ and $\chi=3072$: bare strips, classical edge mean field, a $1\times2$ seam gate, and the full $R+Q$ circuit. Dashed line: previous rigorous bound~\cite{Ovchinnikov1993}; red line: QMC~\cite{Sandvik2026}.
(c) Certified bounds across the XXZ model (blue), compared with QMC~\cite{Caci2024}, coupled-cluster (CCM) results~\cite{BishopLi2017}, and our evaluation of Ovchinnikov's ansatz at each $\Delta$~\cite{Ovchinnikov1993}.}
\label{fig:energy_bound}
\end{figure}

The Heisenberg ground-state energy is known numerically to very high precision, with quantum Monte Carlo giving $\epsilon_0=-0.669441857(7)$~\cite{Sandvik2026} and consistent estimates from variational Monte Carlo~\cite{LiuManousakis1989}, tensor networks~\cite{Corboz2016}, and coupled-cluster calculations~\cite{BishopLi2017}. These calculations, however, do not provide certified upper bounds. By the variational principle, any explicit ansatz gives a rigorous upper bound provided its energy can be evaluated with controlled accuracy. The best previous rigorous bound, $\epsilon_0\leq-0.66068$, was obtained from a fermionic mean-field ansatz~\cite{Ovchinnikov1993}. Another strategy is to solve finite patches and tile the lattice with the resulting patch states~\cite{Oguchi1955,MataArnold1988,Yamamoto2009}. Such product states can be contracted exactly in the thermodynamic limit, but lose quantum correlations across patch boundaries.

Building on this idea, we develop a new certified variational construction that replaces finite patches with infinite strips and restores entanglement across the remaining seams with a shallow quantum circuit, as shown in Fig.~\ref{fig:energy_bound}(a). Writing $\psi$ for the width-$W$ infinite-strip MPS and $U$ for the seam circuit,
\begin{equation}
    |\Psi(\psi,U)\rangle
    =
    U\bigotimes_s|\psi\rangle_s .
    \label{eq:strip_ansatz}
\end{equation}
Ordering the spins column by column reduces every contraction within a
strip to a one-dimensional sequential process, so the reduced density matrix $\rho_A^{(s)}$ of any finite region $A$ on strip $s$ can be obtained directly in the thermodynamic limit. We use four-spin gates $R$ and eight-spin gates $Q$ to restore correlations across the seams, as shown in Fig.~\ref{fig:energy_bound}(a). Because the circuit has finite depth, the backward causal cone of a seam bond $h_{ij}$ has finite support $A\cup B$, with $A\subset s$ and $B\subset s+1$ on the two neighboring strips. Let $U_{ij}$ denote the gates within this causal cone. Since the strips are independent before the circuit is applied,
\begin{equation}
    \langle h_{ij}\rangle_\Psi
    =
    \Tr\!\left[
    \bigl(\rho_A^{(s)}\otimes\rho_B^{(s+1)}\bigr)
    U_{ij}^{\dagger}h_{ij}U_{ij}
    \right].
    \label{eq:causal_cone}
\end{equation}
Thus each bond energy reduces to one-dimensional strip contractions for $\rho_A^{(s)}$ and $\rho_B^{(s+1)}$, followed by a finite contraction over the circuit causal cone.

We optimize the strip state and circuit alternately,
\begin{equation}
\psi_n,U_n
\underset{H_{\rm eff}(U_n)}{\xrightarrow{\rm DMRG}}
\psi_{n+1},U_n
\underset{E(U)}{\xrightarrow{\text{optimize }U}}
\psi_{n+1},U_{n+1}.
\label{eq:variational_optimization}
\end{equation}
For fixed $U$, the circuit defines an effective one-dimensional MPO $H_{\rm eff}(U)$ optimized by DMRG. For fixed $\psi$, the circuit gates are optimized using the finite energy functional $E(U)$. The resulting state is then certified independently. A finite-patch inequality converts its local energy evaluation into a thermodynamic upper bound, with all numerical errors enclosed using verified floating-point bounds and outward-rounded interval arithmetic. The optimization itself does not enter the certificate. Details are given in the End Matter.

Fig.~\ref{fig:energy_bound}(b) shows the resulting hierarchy at the Heisenberg point for $W=12$ and $\chi=3072$. Bare decoupled strips already give $\epsilon_0\leq-0.661161$, surpassing the previous rigorous bound. Adding a classical edge mean field~\cite{MataArnold1988} gives essentially no further improvement, showing that classical boundary polarization alone does not recover the remaining seam energy. In contrast, a $1\times2$ gate acting across each seam bond improves the certified bound to $-0.667203$. The full $R+Q$ circuit gives
\begin{equation}
\epsilon_0\leq-0.668047,
\label{eq:energy_upper}
\end{equation}
only $0.21\%$ above the quantum Monte Carlo energy.

The same construction gives certified upper bounds throughout the XXZ regime used below, as shown in Fig.~\ref{fig:energy_bound}(c). Our variational ansatz preserves the $S^z$ symmetry and is therefore naturally suited to the easy-axis regime $\Delta\geq1$. For $\Delta<1$, where the easy-plane ground state breaks this symmetry, an exact spin rotation maps the energy bound to an easy-axis problem at $\Delta'=2/(1+\Delta)>1$, with the energy rescaled by $(1+\Delta)/2$. The rotation and its certification are described in the End Matter. The resulting bounds closely track quantum Monte Carlo and coupled-cluster benchmarks~\cite{Caci2024,BishopLi2017} and provide the variational energy ceilings used in the bootstrap below.

\textit{Results---}
Our main results are shown in Fig.~\ref{fig:lowerbounds}.
We evaluate the lower bound in Eq.~\eqref{eq:neelbound} on a grid of $\Delta\in[0,2]$, bounding $m_x^2$ for $\Delta\leq 1$ and $m_z^2$ for $\Delta \geq 1$.
Away from $\Delta = 1$, a small operator set ($\sim 2.3\times 10^4$ free moments) with a moderately tight energy ceiling is sufficient to establish order\footnote{The energy ceilings are computed using simpler certified ceilings: finite patches ($8\times 8$ for $\Delta\leq 1$ and $12\times 6$ for $\Delta > 1$), with two-site gates across each seam and energies evaluated in exact rational arithmetic}.
Close to $\Delta = 1$, these bounds are negative, requiring larger operator sets.
With operator sets as many as $1.2\times 10^5$ free moments at $\Delta = 1$, and the tighter variational ceilings described above, we recover a positive bound at every $\Delta$ on the grid.

At $\Delta =0$, the smaller operator set already gives $m_x^2\geq 0.068$, which is about six times larger than the KLS bound $m_x^2 \geq 0.0106$\cite{PhysRevLett.61.2582}.
At the Heisenberg point, our best bound using the largest operator set and the best certified ceiling $\varepsilon_{\rm var}=-0.668047$ is:
\begin{equation}
    m_x^2 \geq 7.63\times 10^{-4}.
\end{equation}
The generalized sum rule uses every correlator $C^X_{\mathbf n}$ that
the operator set can bound: all symmetry-inequivalent separations with
taxicab distance $|\mathbf n|\le6$ (15 shells) for the smaller operator set shown in red in Fig.~\ref{fig:lowerbounds}, and taxicab distance $|\mathbf n|\le16$ (80 shells) for the largest set at $\Delta=1$.
All the bounds are exact certificates, with the certification method described in the End Matter.

We emphasize that a straightforward generalization of the construction also allows us to bootstrap upper bounds on the order parameter without needing the infrared bound\footnote{For any trigonometric polynomial $w(\mathbf{q})\geq 0$ with $w(\mathbf{Q}) = 1$, positivity of $S(\mathbf{q})$ gives $m_x^2\le\int\!\frac{d^2 \mathbf{q}}{(2\pi)^2}\,w(\mathbf q)S(\mathbf q)
=\frac{a_0}{4}+\sum_{\mathbf n}a_{\mathbf n}C^X_{\mathbf n}$, and maximizing
the right-hand side over $\mathcal K$ bounds $m_x^2$ from above. With an
optimized nonnegative $w$ of range 16 and $4.4\times10^4$ free moments we find
$m_x^2\le0.059$ ($m_s\le0.42$), a numerically checked bound.}.
Moreover, we also discuss how to extend the bound certificate at a given $\Delta$ to a finite interval around $\Delta$.

\textit{Acknowledgments}--- NC is grateful to Subir Sachdev for fruitful discussions and for pointing out Ref. \cite{PhysRevB.76.220409}. This work is supported by NSF CAREER grant DMR award No. 2441781 and Alfred P. Sloan Research Fellowship to EK. TW acknowledges additional support from the Harvard Quantum Initiative Fellowship and DOE (USA) office of Basic Sciences Grant No. DE-FG02-03ER46076. MGS is supported by a postdoctoral fellowship from the Julian Schwinger Foundation. PN is supported by the Gordon and
Betty Moore Foundation through Grant GBMF14251 to the
American Chemical Society for the ACS-Moore Postdoctoral Bridge Award in Quantum Materials. This work uses the FASRC Cannon cluster supported by the FAS Division of Science Research Computing Group at Harvard University, the MIT Office of Research Computing and Data (ORCD).
This work used Anvil~\cite{song2022anvil} at Purdue University through allocation PHY260501 from the Advanced Cyberinfrastructure Coordination Ecosystem: Services \& Support (ACCESS) program~\cite{boerner2023access}, which is supported by U.S. National Science Foundation grants \#2138259, \#2138286, \#2138307, \#2137603, and \#2138296. 
The authors acknowledge the use of AI tools Claude Opus 5.5, Fable 5.1 and ChatGPT 6.1 for searching the literature. In addition Claude Opus 5.5 and Fable 5.1 were used for assistance with coding, and algorithmic improvements, particularly the certification procedure.

\bibliography{ref}
\clearpage
\vspace{10em}

\onecolumngrid
\begin{center}
    {\large \textbf{End Matter}}
\end{center}

\renewcommand{\theequation}{A\arabic{equation}}
\setcounter{equation}{0}

\twocolumngrid 
\textit{Reflection positivity}---Consider the reflection $r(x,y)=(1-x,y)$ across the vertical plane $x=1/2$, which divides the even torus into left (L) and right (R) halves with operator algebras $\mathcal{A}_L$ and $\mathcal{A}_R$.
If there is an antilinear isomorphism $\Theta:\mathcal{A}_L\rightarrow\mathcal{A}_R$,
\begin{equation}
 \Theta(cA) = \bar{c}\,\Theta(A),\quad
 \Theta(A^{\dagger}) = \Theta(A)^{\dagger},\quad
 \Theta(AB) = \Theta(A)\Theta(B),
\end{equation}
such that
\begin{equation}
     H = H_L + \Theta(H_L) - \sum_{k\in\partial L} c_kO_k\Theta(O_k), \qquad c_k\geq 0,
\end{equation}
with $H_L,O_k\in\mathcal A_L$. Then $\braket{O\,\Theta(O)}_{\beta} \geq 0$ for all $O\in\mathcal{A}_L$ in every Gibbs state of $H$, and, taking $\beta\rightarrow\infty$, in the ground state~\cite{dyson1978phase,fr6hlich1978phase}. Equivalently, the matrix $\braket{A_a\Theta(A_b)}$ is positive semidefinite, which is the RP constraint we impose.

For the XXZ model, $\Theta$ is the mirror
reflection combined with time reversal,
$\Theta(\mathbf S_i)=-\mathbf S_{r(i)}$, and each bond crossing the plane gives $X_iX_{r(i)}+Y_iY_{r(i)}+\Delta Z_iZ_{r(i)}
=-[X_i\Theta(X_i)+Y_i\Theta(Y_i)+\Delta Z_i\Theta(Z_i)]$, so the $O_k$ are the spin components on the sites next to the plane, with weights $(1,1,\Delta)\ge0$ for every $\Delta\ge0$.

\textit{Deriving the infrared bound.---}
We investigate ordering along the spin component $S$ ($X$ for in-plane and $Z$ for easy-axis order), and $g=1,\Delta$ is the respective coupling in $H$. 
Since neighbours lie on opposite sublattices and $S_i^2=\tfrac14$, each $S$-bond coupling is
$gS_iS_j=\tfrac g2(\tau_i-\tau_j)^2-\tfrac g4$, where $\tau_i = \eta_i S_i$ is the staggered spin. 
We probe the staggered fluctuations with a real site field $h=(h_i)$ which shifts $(\tau_i - \tau_j)\rightarrow (\tau_i -\tau_j - h_i + h_j)$ in the $S$-coupling, with the other two spin components unchanged. 
This gives the perturbed Hamiltonian:
\begin{equation}
  H(h)=H-g\,(\tau,Kh)+\tfrac g2\,(h,Kh),\qquad (Kh)_i=\sum_{j\sim i}(h_i-h_j),
\end{equation}
where $(u,v)=\sum_iu_iv_i, j\sim i$, are the sites coupled to a given $i$. 
Let $Z(h)=\mathrm{Tr}\,e^{-\beta H(h)}$. 
For an unconstrained classical Gaussian field, the shift $\tau\to\tau+h$ would make $Z(h)$ independent of $h$. 
For spins one gets the weaker statement of Gaussian domination,
$Z(h)\le Z(0)$~\cite{frohlich1976infrared,dyson1978phase}, which rests on the following consequence of reflection positivity.

The Hilbert space is a tensor product of the left and right halves, for any $F,G\in\mathcal A_L$, $\mathrm{Tr}[F\,\Theta(G)]=\mathrm{Tr}_L(F)\,\overline{\mathrm{Tr}_L(G)}$,
where $\mathrm{Tr}_L$ is the trace over the left half and the complex
conjugate comes from the antilinearity of $\Theta$. For
operators $A,B,C_k,D_k\in\mathcal A_L$ and coefficients $c_k\ge0$, let
$\mathcal F(A,C;B,D)=\mathrm{Tr}\exp\big[A+\Theta(B)+\sum_kc_kC_k\Theta(D_k)\big]$.
We can expand the exponential as a Dyson series in powers of the crossing terms to express $\mathcal F =
\sum_\alpha \kappa_\alpha\,\mathrm{Tr}_L(F_\alpha)\,\overline{\mathrm{Tr}_L(G_\alpha)}$
with non-negative coefficients $\kappa_{\alpha}$.
$F_\alpha$ is a product of $A$ and
the $C_k$, and $G_\alpha$ is the same product built from $B$ and the $D_k$.
The Cauchy–Schwarz inequality then gives:
\begin{equation}
  |\mathcal F(A,C;B,D)|^2\le\mathcal F(A,C;A,C)\,\mathcal F(B,D;B,D).
  \label{eq:RPZ}
\end{equation}

For $Z(h)$, $A$ is $-\beta$ times the couplings within the left half, $B\in\mathcal A_L$ is such that $\Theta(B) $ is $-\beta$ times the couplings within the right half.
A bond $(i,r(i))$ crossing
the plane, with $i$ on the left, has $S$ term
$\tfrac g2\big(a_i-\Theta(b_i)\big)^2$ with $a_i=\tau_i-h_i$ and
$b_i=\tau_i-h_{r(i)}$, since the reflection exchanges the sublattices, so
$\Theta(\tau_i)=\tau_{r(i)}$, and $h$ is real. Its squares $\tfrac g2a_i^2$
and $\Theta(\tfrac g2b_i^2)$ belong to $A$ and $\Theta(B)$, and its cross term
appears in $-\beta H(h)$ as $\beta g\,a_i\Theta(b_i)$, of the form of
Eq.~\eqref{eq:RPZ} with coefficient $\beta g\ge0$. 
The other two components of the bond cross as $-cO\Theta(O)$ with $c\in\{1,\Delta\}$, as above. 
Setting
$(B,D)=(A,C)$ amounts to replacing $h$ by $h^{(L)}$, defined by
$h^{(L)}_i=h^{(L)}_{r(i)}=h_i$ for $i$ on the left, and similarly for
$h^{(R)}$. 
Thus, we get:
\begin{equation}
\label{eq:zcs}
  Z(h)^2\le Z(h^{(L)})\,Z(h^{(R)}).
\end{equation}
Using translation and rotation symmetry this holds for every mirror plane
between adjacent columns or rows. 
Since constant fields drop out, we may set
$h=0$ at one site; on the remaining fields $(h,Kh)$ grows like $|h|^2$ while
$\|(\tau,Kh)\|$ grows only like $|h|$, so $Z(h)\to0$ as $|h|\to\infty$ and
$Z$ attains its maximum for some $h$. 
But if $h$ is a maximizer, $h^{(L)}$ and $h^{(R)}$ must be maximizer.
By reflecting across different planes in each direction successively eventually leads to the uniform $h$ configuration\cite{bjornberg2022reflection}.
Thus, $Z(0)$ must also be a maximizer and $Z(h)\leq Z(0)$ for every real field configuration $h$.

Now we replace $h$ by $\epsilon h$. 
Since
$f(\epsilon)=\ln Z(\epsilon h)-\ln Z(0)\leq 0$ and $f(0) = 0$, $f'(0)=\beta g\langle(\tau,Kh)\rangle=0$ and $f''(0)\le0$.
Because $H$ and $(\tau,Kh)$ do not commute, the second-order term is a
Duhamel (Kubo) two-point function:
\begin{equation}
  f''(0)=\beta^2g^2\big((\tau,Kh),(\tau,Kh)\big)_{\rm D}-\beta g\,(h,Kh)\le0,
\end{equation}
with
$(U,V)_{\rm D}=Z^{-1}\!\int_0^1ds\,\mathrm{Tr}\big(e^{-(1-s)\beta H}U^\dagger e^{-s\beta H}V\big)$.
For a momentum $\mathbf{k}\neq0$ of the torus, the fields $h_j = \cos(\mathbf{k}\cdot \mathbf{j})$ and
$\sin(\mathbf{k}\cdot \mathbf{j})$ satisfy $Kh=4(1-\gamma_k)h$, where $\gamma_k = \gamma(k)$. 
Applying the inequality to both and adding gives
\begin{equation}
  \beta(\tau_{\mathbf k},\tau_{\mathbf k})_{\rm D}\le\frac{1}{4g(1-\gamma_k)},\qquad
  \tau_{\mathbf k}=\frac{1}{\sqrt {|\Lambda|}}\sum_je^{ik\cdot j}\tau_j=S_{k+Q}.
\end{equation}
Finally, let $\beta\to\infty$ on the fixed torus. The ground state $|0\rangle$
is unique~\cite{marshall1955antiferromagnetism,lieb1962ordering}, and
$\langle0|\tau_{\mathbf{k}}|0\rangle=0$ by spin-rotation symmetry (a $\pi$ rotation
about $z$ for $S=X$, the spin flip for $S=Z$). Hence
$\beta(\tau_{\mathbf k},\tau_{\mathbf k})_{\rm D}$ tends to the static susceptibility
$\chi_S(\mathbf q)=\sum_{n\neq0}\big(|\langle n|S_{\mathbf q}|0\rangle|^2+|\langle n|S_{\mathbf q}^\dagger|0\rangle|^2\big)/(E_n-E_0)$
at $\mathbf q=\mathbf k+\mathbf Q$. Since $\gamma_{\mathbf k}=-\gamma_{\mathbf q}$,
\begin{equation}
  \chi_X(\mathbf{q})\le\frac{1}{4(1+\gamma_q)},\qquad
  \chi_Z(\mathbf{q})\le\frac{1}{4\Delta(1+\gamma_q)},\qquad q\neq Q,
\end{equation}
for all $\Delta\ge0$ ($\Delta>0$ for $\chi_Z$). We use the first bound for
$\Delta\le1$ and the second for $\Delta\ge1$.

\textit{Optimizing the weights}---For fixed cap polynomials $P$, the bound
$\min_{\mathbf y\in\mathcal K}F_{w,P}(\mathbf y)$ is not linear in the weight
coefficients $a=(a_0,a_{\mathbf n})$, because of the positive part
$\{w(\mathbf q)\}_+$. We replace $\{w(\mathbf q)\}_+$ on the momentum grid by
an auxiliary variable $p(\mathbf q)$ with $p(\mathbf q)\ge w(\mathbf q)$ and
$p(\mathbf q)\ge0$. Since $S(\mathbf q)\ge0$ and $S(\mathbf q)\le f_P(\mathbf q)$,
any such $p$ still gives a valid bound, and the tightest choice is
$p=\{w\}_+$. The objective is then linear in $(a,p)$ for fixed $\mathbf y$.
Writing the inner minimization through its dual certificate,
Eq.~\eqref{eq:cert}, every coefficient of that identity is linear in $(a,p)$
and in the dual variables, so maximizing over all of them together, subject to
$w(\mathbf Q)=1$, is a single semidefinite program. Near $\mathbf Q$, where we
keep the square root of the KLS cap, each grid point adds one second-order cone
constraint. Because this program uses a quadrature grid, we use its optimum
only to propose weights; the reported bound comes from the final program at
fixed $w$ and $P$, with the rigorous momentum integrals described below.

\textit{Rigorously bounding the momentum integrals}---For a fixed $w$ and $P$, $F_{w,P}$ is affine in $y$, with Brillouin zone integrals appearing in the constant and the coefficient for each correlator. 
For instance, the coefficient of $G_{2p+1}(\mathbf r)$ is
$-\int\!\frac{d^2 \mathbf{q}}{(2\pi)^2}\,\{w(\mathbf{q})\}_+\,\alpha_{2p+1}(\mathbf{q})\,\gamma_{\mathbf r}(\mathbf{q})$.

During the optimization these integrals are evaluated by quadrature on a momentum grid. 
However, for the final bound this is not enough. 
Quadrature is inaccurate near $\mathbf{Q}$, where $\bar\chi$ diverges. More importantly, to turn a
bound for the quadrature functional into one for the true functional we must
bound their difference for every feasible $\mathbf{y}$.
Using only the norm bounds $|\langle P\rangle|\le1$ this is expensive: the coefficients of $m_5$ multiply long combinations of Pauli expectation values (the expansion
of $\mathrm{ad}_H^5X_0$), so errors of the order of $10^{-9}$ in them cost $\sim10^{-4}$ in the bound. 
We therefore build the final functional directly from rigorous enclosures of the integrals.

First, the cap must hold at every momentum, not only at grid points. 
On each cell where the higher-moment cap is used, the cell's polynomial is applied at every $\mathbf q$ in the cell. 
This is valid because the polynomial is nonnegative on $[0,\infty)$ and $\bar\chi(\mathbf{q})$ and $m_{2p+1}(\mathbf{q})$ are evaluated at the actual $\mathbf q$, not just at the cell's representative. 
Near $\mathbf Q$ we use the Cauchy--Schwarz cap, Eq.~\eqref{eq:CS}, with a tangent point that varies with $\mathbf q$ so that the $1/|\mathbf q-\mathbf Q|$ singularity stays integrable.

Second, each integral is enclosed in an interval. 
On cells where the integrand is analytic we use tensor Gauss--Legendre quadrature with a rigorous remainder
bound from Cauchy estimates. 
Cells on which $w$ changes sign are bisected
repeatedly and bounded crudely at the finest level; they dominate the final
width. 
All arithmetic is rounded outward, and every value of $\sin$ and $\cos$ used, about $10^{10}$ per functional, is checked against correctly rounded multiple-precision evaluation (MPFR), with no failures.

Finally, the coefficients of the functional are set to the midpoints of their intervals, which are exact binary fractions. The half-widths are absorbed through the norm bounds $|\langle P\rangle|\le1$ on the correlators they
multiply. 
For the functionals behind our certified bounds, this enclosure loss, which we denote $\ell$, is
$5$--$7\times10^{-10}$, and the enclosure takes minutes on a single node.

\begin{figure}[t]
\centering
\includegraphics[width=0.75\linewidth]{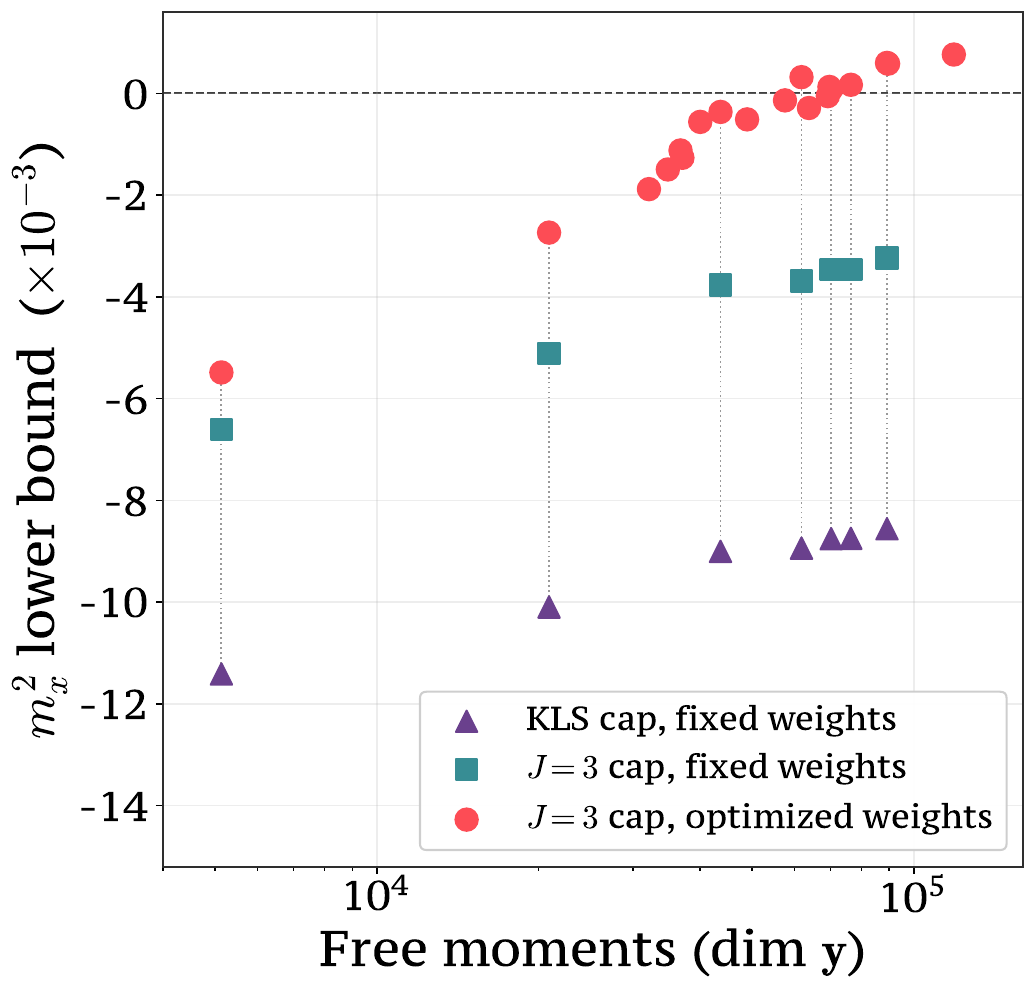}
\caption{\textbf{Variation in lower bounds with the bootstrap program.}
Exact lower bounds on $m_x^2$ at $\Delta=1$ against the number of free moments $\dim\mathbf y$ of the operator set, all with the energy ceiling $-0.668047$. Red circles show the results after optimizing weights and using $J=3$ cap. 
We also plot the respective bounds for a fixed set of weights, with the KLS cap (purple triangles) or the $J=3$ cap (teal squares).
The red points need not increase with $\dim\mathbf y$, because a larger set need not contain a smaller one.
}
\label{fig:opsize}
\end{figure}

\textit{Dependence on the bootstrap program}---\cref{fig:opsize} separates the three choices that enter the bootstrap program at $\Delta=1$: the operator set, the weights $w$, and the cap. 
We built the operator sets by enlarging smaller ones in several independent directions, each aimed at the correlations entering
Eq.~\eqref{eq:neelbound}. 
A larger set therefore need not always contain a
smaller one, and the bound is not monotonic in $\dim\mathbf y$.

Enlarging the operator set alone improves the bound only slowly, by about $3\times10^{-3}$ over the full range shown.
With fixed weights and the KLS cap it stays near $-9\times10^{-3}$. 
Replacing the KLS cap by the $J=3$ cap raises the bound by about $5\times10^{-3}$ for every set considered. 
Optimizing the weights together with the polynomials adds a further $1$–$4\times10^{-3}$, growing with the size of the set, since larger sets bound the longer-range correlations that the optimized weights use. 
This last step is what makes the bound positive.

\textit{Exactly certifying the bootstrap bound}---The semidefinite program is solved using MOSEK \cite{mosek} which only finds an approximate optimum up to numerical tolerances. A rigorous bound $F\ge L$ on $\mathcal K$ requires a certificate of the form
\begin{equation}
  F(\mathbf y)-L=\sum_j\mathrm{tr}\big[Z_j\,A_j(\mathbf y)\big]+\sum_k\mu_k\,g_k(\mathbf y)+\sum_a\nu_a\,e_a(\mathbf y),
  \label{eq:cert}
\end{equation}
holding identically in $\mathbf y$, i.e.\ coefficient by coefficient, with $Z_j\succeq0$ and $\mu_k\ge0$. 
$A_j(\mathbf y)$ are the positivity, perturbative-positivity and reflection-positivity matrices, which depend
affinely on $\mathbf y$ and are required to be positive semidefinite;
$g_k(\mathbf y)\ge0$ are the scalar inequalities (Marshall sign rule, norm
bounds, energy ceiling); and $e_a(\mathbf y)=0$ are the equalities
(normalization, eigenstate conditions, symmetries). 
The first two sums are nonnegative and the last one vanishes on the feasible set, which certifies $F\ge L$.

We note that exact certificates of SDPs have been obtained before using a variety of techniques \cite{Peyrl2008,Jansson2008,Kaltofen2012,Naceur2025,Scheer2026b}. Here, we use a simple strategy that exploits one feature of our SDP: every variable $y_\alpha$ is the expectation value of a Pauli string, so $|y_\alpha|\le1$, and any small residual can therefore be absorbed exactly.

The entire construction of the SDP, including the
elimination of redundant moments and the block diagonalization of the $A_j(\mathbf y)$ by
symmetries, is carried out in exact rational arithmetic; the solver works with a floating-point copy of these data. 
Together with the binary-fraction coefficients of $F$ (see \textit{Rigorously bounding the momentum integrals}) and the energy ceiling entered as an exact decimal, every coefficient in Eq.~\eqref{eq:cert} is an exact rational number.

We factor each approximate
$Z_j\approx B_jB_j^{\mathsf T}$ by setting any slightly negative eigenvalues to 0, round $B_j\rightarrow B_j'$ to binary fractions, and set $Z_j=B_j'B_j'^{\mathsf T}$, which is positive semidefinite by construction irrespective of its rank.
The multipliers $\mu_k$ and $\nu_a$ are rounded in the same way,
with negative $\mu_k$ set to zero. 
After this rounding, Eq.~\eqref{eq:cert} fails only by an affine residual $r_0+\sum_\alpha r_\alpha y_{\alpha}$, which we compute exactly. 
The residual is near machine precision and is removed by absorbing $r_0$ into $L$ and each $r_\alpha y_{\alpha}$ into the norm-bound terms $|r_{\alpha}|(1\pm y_\alpha)$ already present in Eq.~\eqref{eq:cert}.
This gives an exact certificate with a negligible loss, $L_{\rm exact}=L+r_0-\sum_\alpha|r_\alpha|$, without another linear solve or a strictly feasible point. 
For our problems, with up to $1.2\times10^5$ coordinates, this takes minutes on a single node and lowers $L$ by $10^{-13}$ to $10^{-12}$. 
A separate program then re-reads the
exact data and the certificate and re-checks every coefficient of
Eq.~\eqref{eq:cert} in rational arithmetic. 

Combining this with the enclosure loss $\ell$ of the momentum integrals, the ground-state moments $\mathbf y^\star\in\mathcal K$ give $w(\mathbf{Q})\,m_x^2\ge F_{\rm true}(\mathbf y^\star)\ge L_{\rm exact}-\ell$, where $F_{\rm true}$ is the functional with the exact integrals and $w(\mathbf{Q})$ is
an exact rational within $10^{-10}$ of $1$ after rounding. 
The resulting lower bound $(L_{\rm exact}-\ell)/w(\mathbf{Q})$ on $m_x^2$ is an exactly certified rational number.

\textit{Extending the lower bound to an interval around $\Delta$}---Since each certificate at a given $\Delta$ is an exact algebraic identity in which $\Delta$ enters only through explicit rational functions (the stationarity rows, the double commutator and, for $\Delta\ge1$, the infrared bound), the same Gram matrices and multipliers remain a valid certificate at a nearby $\Delta'$, up to a residual that is an explicit rational function of $\Delta'$.
Bounding that residual with the norm constraints $|y_\alpha|\le1$ gives $L(\Delta')\ge L(\Delta)-K|\Delta'-\Delta|$ with an explicit constant $K$, so a single positive certificate in principle establishes order on the whole interval $|\Delta'-\Delta|<L(\Delta)/K$. 
In practice, our certificates have large $K$, leading to narrow intervals of width $\sim 0.002$, which would require a denser sampling of the $\Delta$ grid to certify order for the entire range.
In this work, we only report bounds at the computed values of $\Delta$.

\textit{Certifiable strip construction}---The purpose of this step is to turn the numerically optimized strip MPS into an exact variational quantum state. We choose a sequential ordering of the spins along the strip. At step $i$, the real MPS tensor $V_i$ takes an incoming auxiliary state, emits physical spin $i$, and produces the outgoing auxiliary degree of freedom. For an exactly normalized MPS tensor one would have $V_i^T V_i=I$. The finite-precision tensor is only approximately isometric, so we rigorously certify $\|V_i^T V_i-I\|_{\rm op}\leq\beta_i$. We choose $a_i=(1+\beta_i)^{-1}$, which guarantees $a_iV_i^TV_i\leq I$, and define $D_i=I-a_iV_i^TV_i\succeq0$. The certified one-site preparation map is
\begin{equation}
\mathcal F_i(\sigma_i)
=
a_iV_i\sigma_iV_i^T
+\Tr(D_i\sigma_i)|r_i\rangle\langle r_i| ,
\label{eq:certified_channel}
\end{equation}
where $\sigma_i$ is the incoming auxiliary density matrix and $|r_i\rangle$ is a fixed normalized state of the emitted spin and outgoing auxiliary space. The first term follows the optimized MPS tensor, while the second restores the probability removed by the rescaling. Since $D_i\succeq0$ and $a_iV_i^TV_i+D_i=I$, $\mathcal F_i$ is completely positive and trace preserving (CPTP). The outgoing auxiliary state is $\sigma_{i+1}=\Tr_i\mathcal F_i(\sigma_i)$, where the trace is over the emitted spin. Repeated application of the $\mathcal F_i$, with each subsequent map acting only on the outgoing auxiliary degree of freedom, therefore defines an exact positive normalized many-body state on the strip. No fixed-point assumption is required.

\textit{Seam circuit}---A product of certified strips is already a valid two-dimensional variational state, but it contains no entanglement across neighboring strips. We restore these correlations with a finite-depth unitary circuit across each seam. The simplest circuit in Fig.~\ref{fig:energy_bound}(b) applies one two-spin gate to each bond crossing a seam. Reality and conservation of total $S^z$ leave the generator $A=\theta(S_i^+S_j^- - S_i^-S_j^+)$, where $S^\pm=X\pm iY$. We parameterize the gate by
\begin{equation}
U(A)=(I+A)^{-1}(I-A).
\label{eq:seam_gate}
\end{equation}
Since $A$ is real and antisymmetric, $U(A)$ is exactly orthogonal. Since $A$ preserves total $S^z$, so does the gate. The full circuit in Fig.~\ref{fig:energy_bound}(a) uses the same construction with larger gates. A four-spin gate $R$ acts on a $2\times2$ region crossing the seam, followed by an eight-spin gate $Q$ on a $2\times4$ region, with the pattern repeated every two columns. Their generators are real, antisymmetric, and block diagonal in total $S^z$. The resulting circuit therefore preserves the $U(1)$ symmetry of the strip source. Because the circuit has finite depth, propagating any bond Hamiltonian backward through it produces a finite causal cone involving at most two neighboring strips.

\textit{Thermodynamic energy certificate}---The remaining issue is that the numerical infinite-MPS environment need not be an exact fixed point of the certified strip channel. Let $T$ be the auxiliary CPTP channel obtained by composing the certified preparation maps over one two-column cell. We define an exact reference auxiliary state directly from the stored Schmidt coefficients,
\begin{equation}
\sigma_0=
\frac{\operatorname{diag}(S_i^2)}
{\sum_jS_j^2}.
\label{eq:reference_source}
\end{equation}
The stored $S_i$ are treated as exact dyadic numbers, so $\sigma_0\succeq0$ and $\Tr\sigma_0=1$ hold algebraically.

To certify stationarity without subtracting two nearly equal $O(1)$ density matrices, we propagate the stationarity defect itself. Define
\begin{equation}
D_0=T(\sigma_0)-\sigma_0,
\qquad
D_k=T^k(D_0).
\label{eq:defect_propagation}
\end{equation}
Then
\[
D_k=T^{k+1}(\sigma_0)-T^k(\sigma_0).
\]
For a chosen burn-in depth $n_{\rm b}$, we set $\sigma=T^{n_{\rm b}}(\sigma_0)$, so
\begin{equation}
T(\sigma)-\sigma=D_{n_{\rm b}},
\qquad
\sigma-\sigma_0=\sum_{k=0}^{n_{\rm b}-1}D_k .
\end{equation}
Verified propagation therefore gives
\begin{equation}
\delta\geq\|D_{n_{\rm b}}\|_1,
\qquad
d_0\geq\sum_{k=0}^{n_{\rm b}-1}\|D_k\|_1
\geq\|\sigma-\sigma_0\|_1 .
\label{eq:source_certificate}
\end{equation}
The small defect $D_k$, rather than an $O(1)$ density matrix, is propagated in round-to-nearest binary64 with explicit roundoff majorants, and all resulting bounds are enclosed outward using Arb ball arithmetic. Thus the certified residual is not limited by the $O(10^{-16})$ absolute scale associated with subtracting two normalized density matrices. For any nonnegative integer $k$, trace-norm contractivity of the CPTP channel and a telescoping sum give
\begin{equation}
\|T^k(\sigma)-\sigma\|_1\leq k\delta .
\label{eq:telescoping_source}
\end{equation}

The exact variational state used for the upper bound is obtained by initializing each strip patch with $\sigma$, applying the certified channel for $L$ two-column cells, periodically repeating the resulting patches, and then applying the seam circuit. This is an exact positive normalized many-body state for every finite $L$. The patch can be extremely long, but Eq.~\eqref{eq:telescoping_source} allows its bulk energy to be bounded without contracting all $L$ cells individually.

We evaluate the reference energy using the exact diagonal state $\sigma_0$. For every physical bond in one two-column circuit unit cell, the bond Hamiltonian is propagated backward through the seam circuit to its finite causal cone and contracted using $\sigma_0$ and the certified strip maps. Every physical bond is included exactly once. Each bond expectation value is evaluated with rigorous floating-point error bounds, and the resulting intervals are accumulated using outward-rounded arithmetic. Normalizing the total bond-energy upper bound by the $2W$ sites in the unit cell defines the certified reference energy density $e_{\rm ref}$.

It remains to compare this reference contraction with the exact length-$L$ patch. For one strip, an auxiliary state entering a bulk causal cone after $k$ cells differs from $\sigma_0$ by at most
\begin{equation}
\|T^k(\sigma)-\sigma_0\|_1
\leq d_0+k\delta .
\end{equation}
Every bond causal cone involves at most two strips, so the standard tensor-product trace-distance bound gives an additional factor of two. Longitudinally, the $R+Q$ causal cone extends at most two two-column cells beyond the cell containing the bond. Including one additional cell for full-cell coverage, together with the anchoring and closing applications of $T$, every auxiliary state entering a bulk causal cone occurs within at most $L+5$ applications of $T$. The source density matrix of any bulk causal cone therefore differs from the reference source by at most
\begin{equation}
2\bigl[d_0+(L+5)\delta\bigr]
\label{eq:source_error}
\end{equation}
in trace norm.

For a spin-$1/2$ Heisenberg bond $h=\mathbf S_i\cdot\mathbf S_j$, the spectrum is $\{1/4,-3/4\}$, so $\|h\|_{\rm op}=3/4$. Unitary evolution through the seam circuit preserves trace distance, partial trace cannot increase it, and H\"older's inequality gives
\[
|\Tr(h\,\delta\rho)|
\leq
\|h\|_{\rm op}\|\delta\rho\|_1 .
\]
Since there are two bonds per site on the square lattice, Eq.~\eqref{eq:source_error} changes the bulk energy density by at most
\begin{equation}
3\bigl[d_0+(L+5)\delta\bigr].
\end{equation}

The bulk comparison excludes at most four cells at each end of a patch, giving an uncontrolled fraction $8/L$. The Heisenberg bond spectrum bounds the energy per site to $[-3/2,1/2]$, whose width is $2$, so these boundary regions contribute at most $16/L$. Applying the variational principle to the exact periodic patch state therefore gives
\begin{equation}
\epsilon_0
\leq
e_{\rm ref}
+3d_0
+3(L+5)\delta
+\frac{16}{L}.
\label{eq:finite_patch_certificate}
\end{equation}
Thus $e_{\rm ref}$ is evaluated entirely at the exact reference state $\sigma_0$, $d_0$ connects this reference contraction to the propagated state $\sigma$ used to initialize the patch, and $\delta$ controls the subsequent drift across the patch.

After $e_{\rm ref}$, $\delta$, and $d_0$ have been certified, $L$ is chosen to minimize Eq.~\eqref{eq:finite_patch_certificate}. Short patches incur a boundary cost proportional to $1/L$, while long patches accumulate a stationarity error proportional to $L\delta$. Minimizing the $L$-dependent terms gives
\begin{equation}
L^\star=
\sqrt{\frac{16}{3\delta}} .
\label{eq:certificate_length}
\end{equation}
For the isotropic certificate used in Eq.~\eqref{eq:energy_upper}, $n_{\rm b}=11$ gives
\begin{equation}
\delta\leq1.684\times10^{-19},
\qquad
d_0\leq9.01\times10^{-12}.
\end{equation}
The corresponding optimal patch length is $L\simeq5.63\times10^9$ cells. The patch correction,
\[
3(L+5)\delta+\frac{16}{L},
\]
is at most $5.69\times10^{-9}$, while the reference-state shift contributes at most $3d_0\leq2.71\times10^{-11}$. The total correction in Eq.~\eqref{eq:finite_patch_certificate} is therefore at most $5.72\times10^{-9}$. Related bootstrap approaches instead bound tensor-network contractions by constraining the space of admissible MPS environments~\cite{OnoZhangPo2026}. For the XXZ calculations, the corresponding norm allowances are multiplied by the conservative factor $\max(1,\Delta)$ used in the certificate.

\textit{Easy-plane mapping}---The strip source and seam circuit preserve total $S^z$, which is naturally suited to the easy-axis regime. For $\Delta<1$, rather than introduce a separate symmetry-breaking variational family, we map the easy-plane energy problem exactly to an easy-axis one. Define the global $\pi/2$ rotations $R_x=\exp[-i(\pi/2)\sum_jX_j]$ and $R_y=\exp[-i(\pi/2)\sum_jY_j]$. They obey
\begin{equation}
\frac12\left[
R_x^\dagger H(\Delta)R_x
+
R_y^\dagger H(\Delta)R_y
\right]
=
\frac{1+\Delta}{2}
H\!\left(\frac{2}{1+\Delta}\right).
\label{eq:easy_plane_rotation}
\end{equation}
For $\Delta<1$, the transformed coupling $\Delta'=2/(1+\Delta)$ lies in the easy-axis regime. If $\rho'$ is any certified variational state for $H(\Delta')$, then
\begin{equation}
\rho=
\frac12\left(
R_x\rho'R_x^\dagger
+
R_y\rho'R_y^\dagger
\right)
\end{equation}
is again a positive normalized state, and Eq.~\eqref{eq:easy_plane_rotation} gives
\begin{equation}
\Tr[\rho H(\Delta)]
=
\frac{1+\Delta}{2}
\Tr[\rho'H(\Delta')] .
\label{eq:easy_plane_energy}
\end{equation}
The variational principle therefore gives
\begin{equation}
\epsilon_0(\Delta)
\leq
\frac{1+\Delta}{2}\,
\varepsilon_{\rm var}(\Delta'),
\label{eq:easy_plane_bound}
\end{equation}
which gives the easy-plane bounds shown in Fig.~\ref{fig:energy_bound}(c).

\end{document}